\documentclass[prc,superscriptaddress,unsortedaddress,twocolumn]{revtex4}
\usepackage{graphicx,epsf,bm,amsmath,color}
\usepackage{here}
\def\bK{\mbox{\boldmath $K$}}
\def\bk{\mbox{\boldmath $k$}}

\def\br{\mbox{\boldmath $r$}}
\def\bR{\mbox{\boldmath $R$}}
\def\bs{\mbox{\boldmath $s$}}

\begin{document}
\title{$\Xi$ hyperons in the nuclear medium described by chiral NLO interactions}
\author{M. Kohno}
\affiliation{Research Center for Nuclear Physics, Osaka University, Ibaraki 567-0047,
Japan}

\begin{abstract}
Properties of the baryon-baryon interactions in the strangeness $S=-2$ sector
of chiral effective field theory at the next-to-leading order (NLO) level are
explored by calculating $\Xi$ single-particle potentials in symmetric nuclear matter.
The results are transformed to the $\Xi$ potential in finite nuclei by a local-density
approximation with convolution by a Gaussian form factor to simulate finite-range
effects. The $\Xi$ potential is repulsive in a central region, and attractive in a
surface area when the $\Xi$ energy is low. The attractive pocket can lower
the $\Xi^-$ $s$ and $p$ atomic states. The obtained binding energies in $^{12}$C
and $^{14}$N are found to be conformable with those found in emulsion experiments
at Japan's National Laboratory for High Energy Physiks (KEK). $K^+$ spectra
of $(K^-, K^+)$ $\Xi$ production inclusive processes on $^9$Be and $^{12}$C are also
evaluated, using a semi-classical distorted wave method. The absolute values
of the cross section are properly reproduced for $^9$Be, but the peak locates at a
lower energy position than that of the experimental data. The calculated
spectrum of $^{12}$C should be compared with the forthcoming result from the
new experiments recently carried out at KEK with better resolution than before.
The comparison would be valuable to improve the understanding of the $\Xi N$
interaction, the parametrization of which has still large uncertainties.
\end{abstract}

\maketitle

\section{Introduction}
The realistic description of interactions between baryons is basic for the microscopic
study of hadron many-body systems. In the last two decades, a new framework for the
explicit derivation of the interactions has been developed in chiral effective field theory
(ChEFT) \cite{EHM09, ME11}. In the nonstrangeness sector, the ChEFT interactions
are now employed as standard inputs for ab initio calculations of nuclear structures
and reactions. The parametrization of the baryon-baryon interactions in the strangeness
sectors has also progressed, although experimental information to determine
the coupling constants is not sufficient.

Leading order (LO) ChEFT interactions in the strangeness $S=-1$ and $S=-2$ sectors were
constructed by Polinder, Haidenbauer, and Mei{\ss}ner \cite{POL06}. Then, the extension
to the next-to-leading order (NLO) was made by Haidenbauer \textit{et al.} \cite{HAID13}
for $\Lambda N$ and $\Sigma N$ interactions, and by Haidenbauer, Mei{\ss}ner, and
Petschauer \cite{HAID16} for $\Lambda\Lambda$, $\Lambda\Sigma$, $\Sigma\Sigma$,
and $\Xi N$ interactions. Leading three-baryon interactions were also formulated
by Petschauer \textit{et al.} \cite{PET16}.

The $\Lambda N$ interaction is fairly well documented from by the experimental data of
$\Lambda$ hypernuclei. The $\Lambda N$ interaction which accounts for the $\Lambda$
binding energies in hypernuclei tends to predict deeper binding in high-density nuclear
and neutron matter. Therefore, $\Lambda$ hyperons should appear at some density,
$2-3$ times the normal density, to bypass the increasing neutron Fermi energy
in neutron star matter, which inevitably softens the equation of state (EOS) of the
high-density matter. This scenario is not favorable even for standard neutron
stars with the mass of about $1.4$ solar mass. The recent observation of massive
neutron stars with about twice the solar mass has made the situation more uncertain.
This difficulty is called the hyperon puzzle. It is naturally possible to introduce ad
hoc two-body and/or three-body repulsive forces to make the EOS hard, and
various models have been proposed. On the other hand, it is shown \cite{HMKW17,Koh17}
that the NLO $\Lambda N$ interaction predicts
shallow $\Lambda$ binding above normal density, which implies that the $\Lambda$
hyperon does not become energetically favored in high-density neutron star matter.
Therefore, the description by the ChEFT interaction offers a possibility to resolve
the hyperon puzzle.

Because experimental scattering data in the $S=-2$ sector are much scarcer than
in the $S=-1$ sector, the present parametrization includes many uncertainties.
Regarding the recent experimental situation for the $\Xi$ states in nuclei, a few possible
bound states have been found in nuclear emulsion experiments at Japan's National
Laboratory for High Energy Physiks (KEK).
Nakazawa \textit{et al.} \cite{Naka15} reported the first evidence of a deeply bound
state of the $\Xi^-$-$^{14}$N system. Shallow $\Xi^-$-$^{12}$C bound states
are also plausible \cite{Aok09}. The production of the $\Xi^-$ hyperon in
$(K^-, K^+)$ reactions on nuclei is another source of information on the $\Xi^-$-
nucleus interaction. The $K^+$ spectra of the $(K^-,K^+)$ $\Xi^-$ production inclusive
reactions on $^{12}$C were measured in the past at KEK \cite{Fuk98} and BNL \cite{Kha00}.
No peak structure revealing $\Xi^-$ bound states was found. Nevertheless,
the analysis of the spectrum \cite{Kha00} suggested that the $\Xi$ potential
is attractive with the depth of about 14 MeV in a standard Woods-Saxon form. This value
has been accepted as a canonical attractive strength, though an independent
analysis \cite{KH10} showed that the $\Xi$ potential could be very weak.
Recently, KEK conducted new $(K^-, K^+)$ $\Xi^-$ production reactions on $^{12}$C
\cite{Nag18} with better accuracy than before, and the final result of the analysis is awaited.

On the basis of this experimental progress, it is interesting to explore the properties
of the $\Xi N$ interactions in the nuclear medium and the $\Xi$-nucleus potentials,
which are described by the ChEFT interactions in the strangeness $S=-2$ sector,
and compare the results with the experimental data available at present and expected
in the near future. Such an exploration elucidates the character of the present 
description in the $S=-2$ sector and could help to improve the parametrization.
Because $\Xi^-$ hyperons have a possibility to appear at high density and can
influence the EOS of neutron star matter, a better understanding of the interactions
in the $S=-2$ sector is also relevant for the study of neutron star matter.

In this article, the framework of lowest-order Bruckner theory (LOBT) is used to
obtain $\Xi N$ interactions in the nuclear medium. The Bethe-Goldstone, or $G$-matrix,
equation in the LOBT takes care of the strong short-range or high-momentum
component of the bare interaction. The coupling of the $\Xi N$ state with other
baryon channels such as $\Lambda\Lambda$, $\Lambda\Sigma$, and $\Sigma\Sigma$
is incorporated nonperturbatively by the baryon-channel coupled $G$-matrix equation.
In order to solve the baryon-channel coupling in the nuclear medium, single-particle
potentials of the $\Lambda$ and $\Sigma$ hyperons as well as the nucleon have to be
prepared beforehand. The present study is based on the preceding calculations of
properties of nuclear matter \cite{Koh13} and the successive investigation of the
$\Lambda$ and $\Sigma$ potentials in nuclear matter \cite{Koh17}, using the
baryon-baryon interactions of ChEFT including effects of two-pion exchange
three-baryon forces \cite{PET16}.

The parameters of the chiral NLO interactions in the strangeness $S=-2$ sector
were recently updated \cite{HAID19} by considering the recent experimental
evidence of a $\Xi^-$ bound state in $^{14}$N. In the present calculations, these
parameters with a cutoff scale of 550 MeV are employed.

In Sec. II, $\Xi$ single-particle potentials are evaluated in symmetric nuclear matter (SNM).
The potentials obtained are transformed, in Sec. III, to $\Xi$ potentials in finite nuclei by
a local-density approximation (LDA). To simulate finite range effects which are absent
in the LDA, a Gaussian form-factor is convoluted with the range of 1 fm. The
energy-dependent potential thus obtained is parametrized in a convenient functional form
in Sec. IV, and $\Xi^-$ bound states predicted by the parametrized $\Xi$ potential
are discussed for $^{12}$C and $^{14}$N. $K^+$ spectra of $(K^-,K^+)$ $\Xi$
production inclusive processes on $^9$Be and $^{12}$C are calculated in Sec. V,
using a semiclassical distorted-wave method. The conclusion follows in Sec. VI.

\section{$\Xi$ potential in symmetric nuclear matter}
The standard method to consider baryon-baryon interactions in the nuclear medium
is the Bethe-Goldstone, or $G$-matrix, equation in Brueckner theory.
The singular short-range or high-momentum component of the bare interaction $V$
is properly regularized by the $G$-matrix equation:
\begin{equation}
 G=V+V\frac{Q}{\omega-H_0}G,
\end{equation}
in which medium effects due to the Pauli principle and the change of baryon
propagators are taken into account by the exclusion operator $Q$ and the potential
insertion in the Hamiltonian $H_0$. The reaction matrix $G$ depends on the starting energy
$\omega$, which is a sum of single-particle energies of the interacting two baryons.

The $\Xi N$ state couples to $\Lambda\Lambda$, $\Lambda\Sigma$, and $\Sigma\Sigma$
channels, depending on the spin, isospin, and orbital angular momentum.
The coupling effect is taken care of in a nonperturbative manner by the baryon-channel
coupled  $G$-matrix equation. The single-particle potentials of the $\Lambda$ and
the $\Sigma$ hyperons as well as the nucleon, which are included in $H_0$, are to be
prepared as input.

The present author \cite{Koh13} has discussed nuclear matter saturation properties in
the lowest-order Brueckner theory, using the next-to-next-to-next-leading order NN
forces \cite{EGM05} and the next-to-next-leading-order three-nucleon forces (3NFs)
\cite{NLO3NF} in ChEFT. The important repulsive effects for bringing about the
nuclear saturation properties from the 3NFs which are fixed by the coupling constants
in the NN sector were demonstrated. The remaining two parameters of the 3NFs
which are not determined in the NN sector were adjusted to reproduce the empirical
saturation properties reasonably well. On the basis of these nuclear matter calculations,
the NLO hyperon-nucleon interactions derived by the J\"{u}lich-Bonn-M\"{u}nchen
group \cite{HAID13} were used to evaluate $\Lambda$ and $\Sigma$ single-particle
potentials in SNM and in pure neutron matter \cite{Koh17}. The contributions of
the two-pion exchange $\Lambda NN$ and $\Lambda NN$-$\Sigma NN$
forces \cite{PET16} were also considered. The resulting hyperon potentials
are reasonable in view of the empirical properties of hypernuclei.

These N, $\Lambda$, and $\Sigma$ single-particle potentials obtained in the previous
studies are used in the propagator of the present $G$-matrix equation for the $\Xi$
hyperon. The self-consistent $\Xi$ single-particle potential is determined in an iterative
way. It is not necessary to explain the details of the calculation because $G$-matrix
calculations are basically standard. Here, only the key points are noted.
(1) The continuous choice is used for the intermediate spectra.
An effective mass approximation is not used, but single-particle energies are interpolated
from the values at the mesh points in momentum space.
(2) The angle-average approximation is introduced for the Pauli operator in the
numerator and the energies in the denominator of the propagator.
(3) Partial waves up to the total angular momentum $J=6$ are included.
(4) The cutoff factor $e^{-(k^2/(2\Lambda^2))^3}$ is introduced to the single-particle
potential in the propagator in the $G$-matrix equation. The cutoff scale $\Lambda=550$
MeV is the same as the scale of the NLO interactions in the $S=-2$ sector \cite{HAID13}.
This procedure stabilizes the iteration to achieve the self-consistent results of
baryon single-particle potentials, but hardly change the potentials at low momentum.

\begin{figure}[bht]
\centering
 \includegraphics[width=0.4\textwidth,clip]{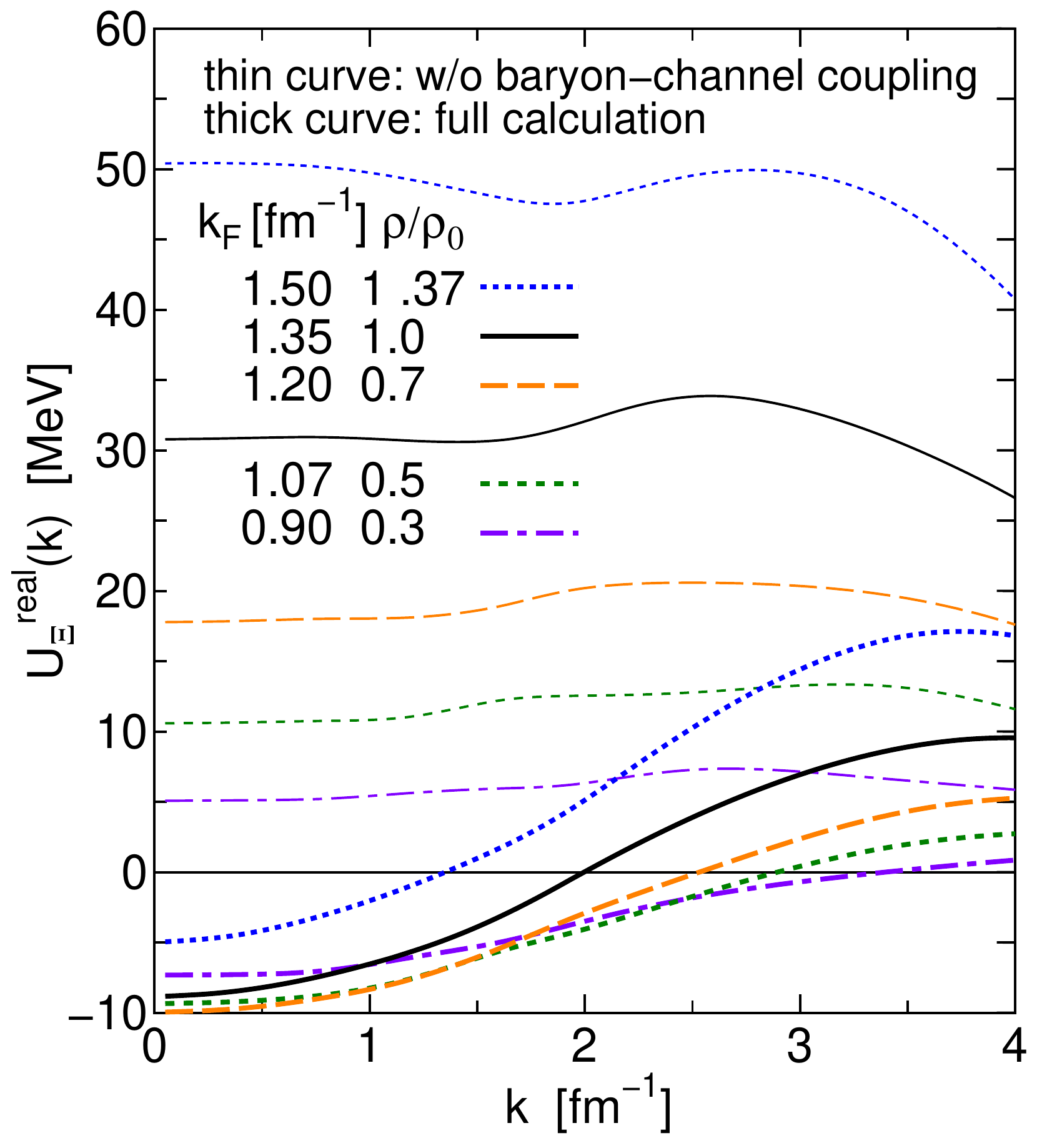}
 \includegraphics[width=0.4\textwidth,clip]{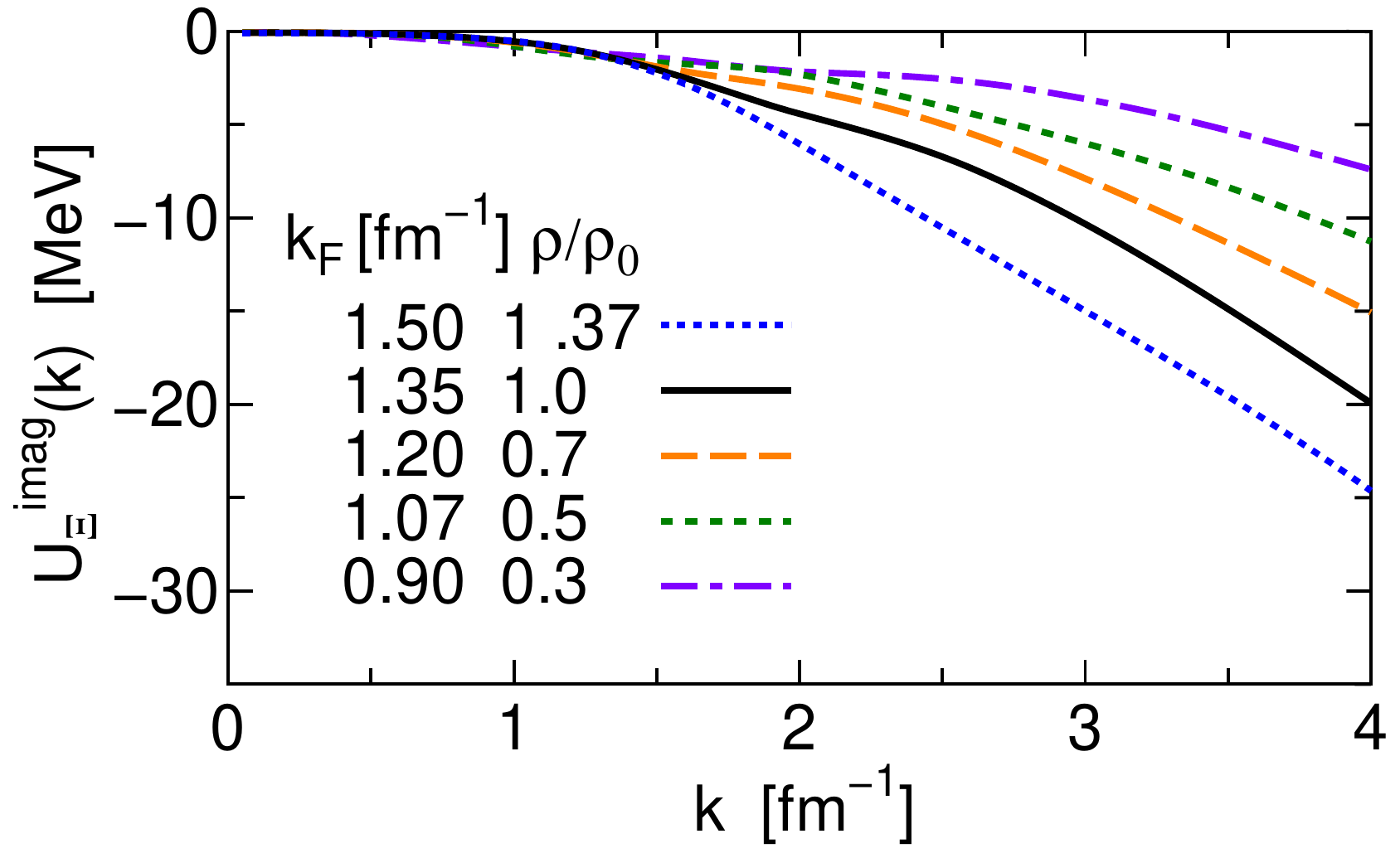}
 \caption{Momentum dependence of the real and imaginary parts of $\Xi$ single-particle
potentials in SNM at various Fermi momenta $k_F$, using the ChEFT NLO interactions
in the strangeness $S=-2$ sector \cite{HAID13}. $\rho/\rho_0$ is the ratio of the
density $\rho=2k_F^3/(3\pi^2)$ to the normal density $\rho_0$ at $k_F=1.35$ fm$^{-1}$.
The thick and thin curves in the real part represent the results with and without
baryon-channel coupling effects.}
\label{fig:sppt}
\end{figure}

Calculated results in SNM as a function of the momentum $k$ are presented in
Fig. \ref{fig:sppt} up to $k=4$ fm$^{-1}$. For the real part, results of the calculations with
switching off the baryon-channel coupling are also shown. The potential beyond
$k=4$ fm$^{-1}$ decreases because of the cutoff scale of 550 MeV. Without the coupling
the $\Xi$ potential is repulsive and its strength increases with growing the density.
Because there is no strangeness -2 meson exchange and no non-local effect from the
baryon-channel coupling, the $U_{\Xi}^{real}(k)$ without the coupling is flat as a function
of the momentum $k$. The weakly attractive $\Xi$ potential at low momentum is seen
to be generated by the baryon-channel coupling. The $k$ dependence of the $\Xi$
potential is not large. The value of $U_{\Xi}^{real}(k=0)=-8.8$
MeV at $k_F=1.35$ fm$^{-1}$ is slightly more attractive than the corresponding value of
$-3.8$ MeV given by Haidenbauer and Mei{\ss}ner \cite{HAID19} with the same interaction,
which is reasonable because of the continuous prescription for the intermediate spectra
of $G$-matrix equations in contrast to the gap choice in Ref. \cite{HAID19}.
It is noted that other baryon-baryon interactions, such as the Kyoto-Niigata quark-model
potential fss2 \cite{FSN07} and the recent Nijmegen ESC08c \cite{ESC08C}, also predict rather
shallow attractive $\Xi$ single-particle potential in the nuclear medium \cite{Koh09,ESC08C}

The imaginary part, which indicates the strength of creating nucleon particle-hole excitations
and/or the transition to the $\Lambda\Lambda$ channel is seen to be small particularly
at low energies, which is common in the calculations in  \cite{Koh09,YMR10}.
The imaginary potential is related to the broadening of the $K^+$ spectrum of $(K^-,K^+)$
$\Xi^-$ production inclusive reactions, which is discussed in Sec. IV.
  
\begin{figure}[bht]
\centering
 \includegraphics[width=0.4\textwidth,clip]{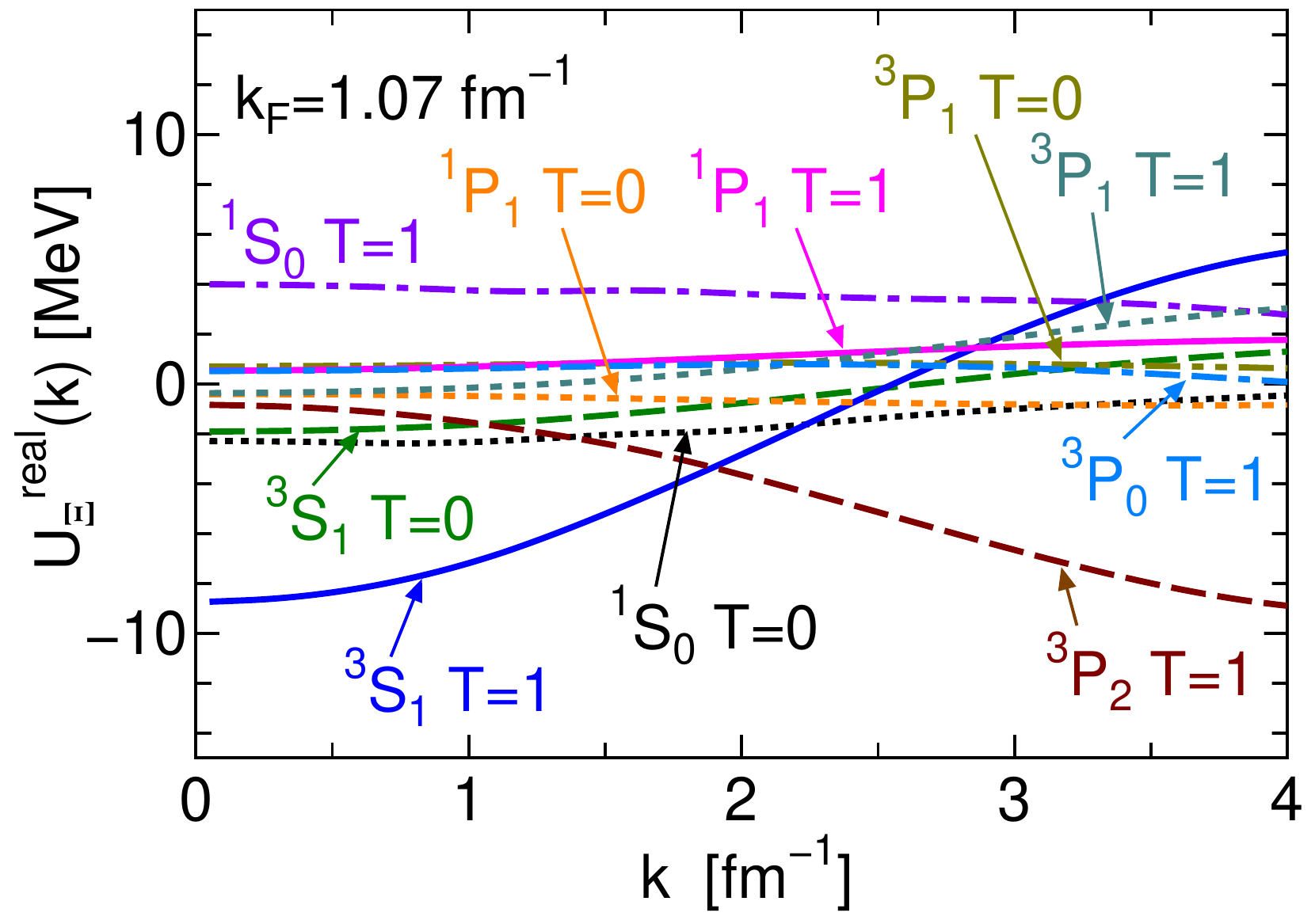}
 \includegraphics[width=0.4\textwidth,clip]{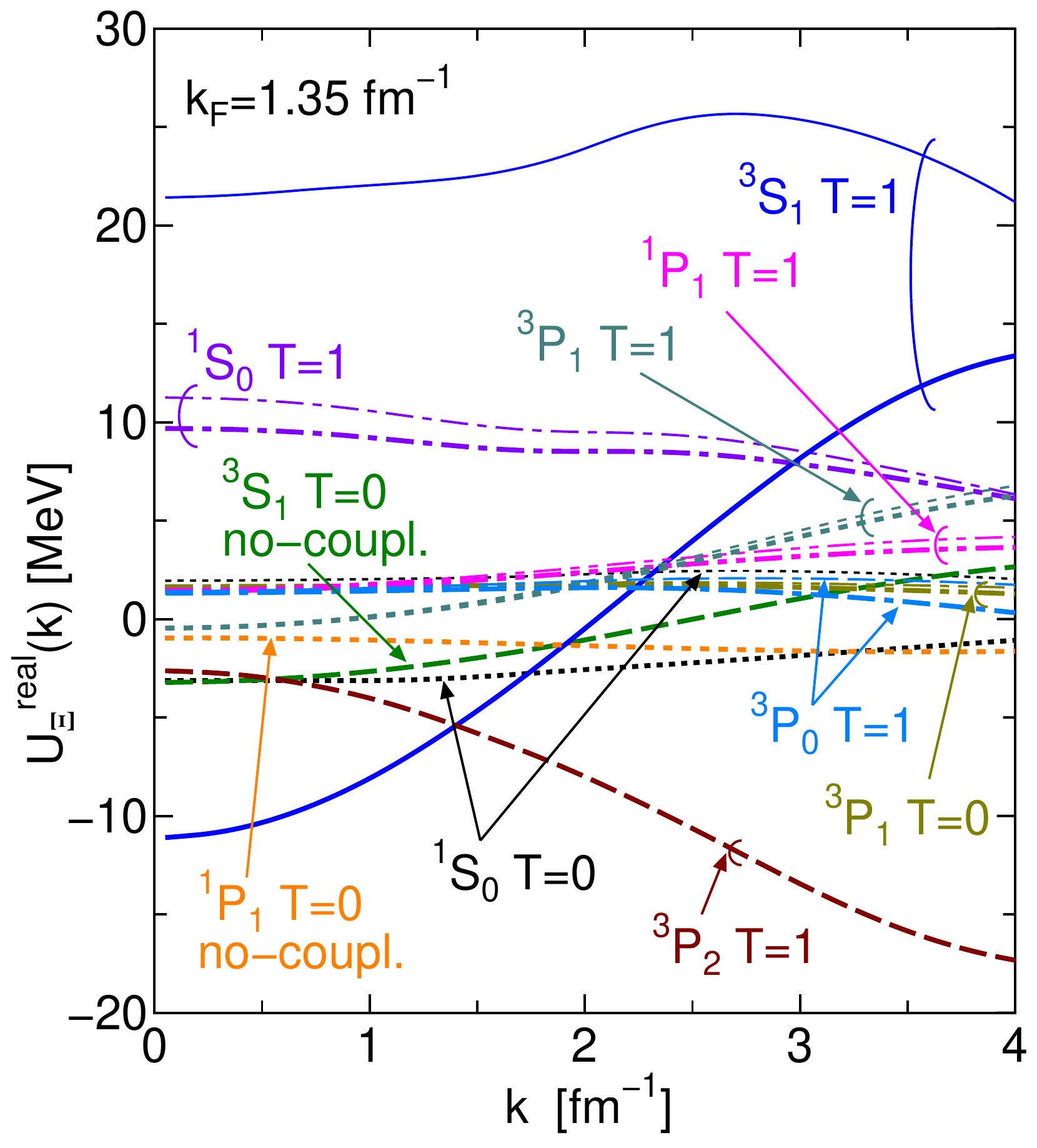}
 \caption{Partial wave contributions to the real part of $\Xi$ single-particle potentials in
SNM at $k_F=1.07$ fm$^{-1}$ and $k_F=1.35$ fm$^{-1}$. The thick curves denote the results
of the full calculations. The thin curves in the lower panel represent the results
with switched off baryon-channel coupling. Note that the $^3$S$_1$
$T=0$ and $^1$P$_1$ $T=0$ states have no coupling to other baryon channels.
The thick and thin curves are indistinguishable in some partial waves.}
\label{fig:sppw}
\end{figure}

The partial-wave contributions to the $\Xi$ single-particle potential and their
$k_F$-dependence are useful to elucidate the properties of the $\Xi N$ interactions
under consideration. Figure \ref{fig:sppw} shows each $^{2S+1}L_J$-state contribution with
a total spin $S$, an orbital angular momentum $L$, and a total spin $J$ for $L=0$
(denoted by S) and $L=1$ (denoted by P) to the real part at the Fermi momenta $k_F=1.07$
and $1.35$ fm$^{-1}$. The effect of the baryon-channel coupling is also presented
in Fig. \ref{fig:sppw} for $k_F=1.35$ fm$^{-1}$, showing the results by thin curves when
the baryon-channel coupling is switched off. The coupling effect is particularly large
in the $T=1$ $^3$S$_1$ $\Xi N$-$\Lambda\Sigma$-$\Sigma\Sigma$ channel, which
brings about an attractive contribution at low momentum. The $T=0$ $^3$S$_1$ state,
which has no baryon-channel coupling, provides weak attraction. The $T=0$ $^1$S$_0$
contribution changes its sign from positive to negative by the coupling. The $T=1$ $^1$S$_0$
state gives a dominant repulsive contribution, and on the other hand the $T=1$ $^3$P$_2$
state provides a large attractive contribution especially at large momentum.

\begin{figure}[bht]
\centering
 \includegraphics[width=0.4\textwidth,clip]{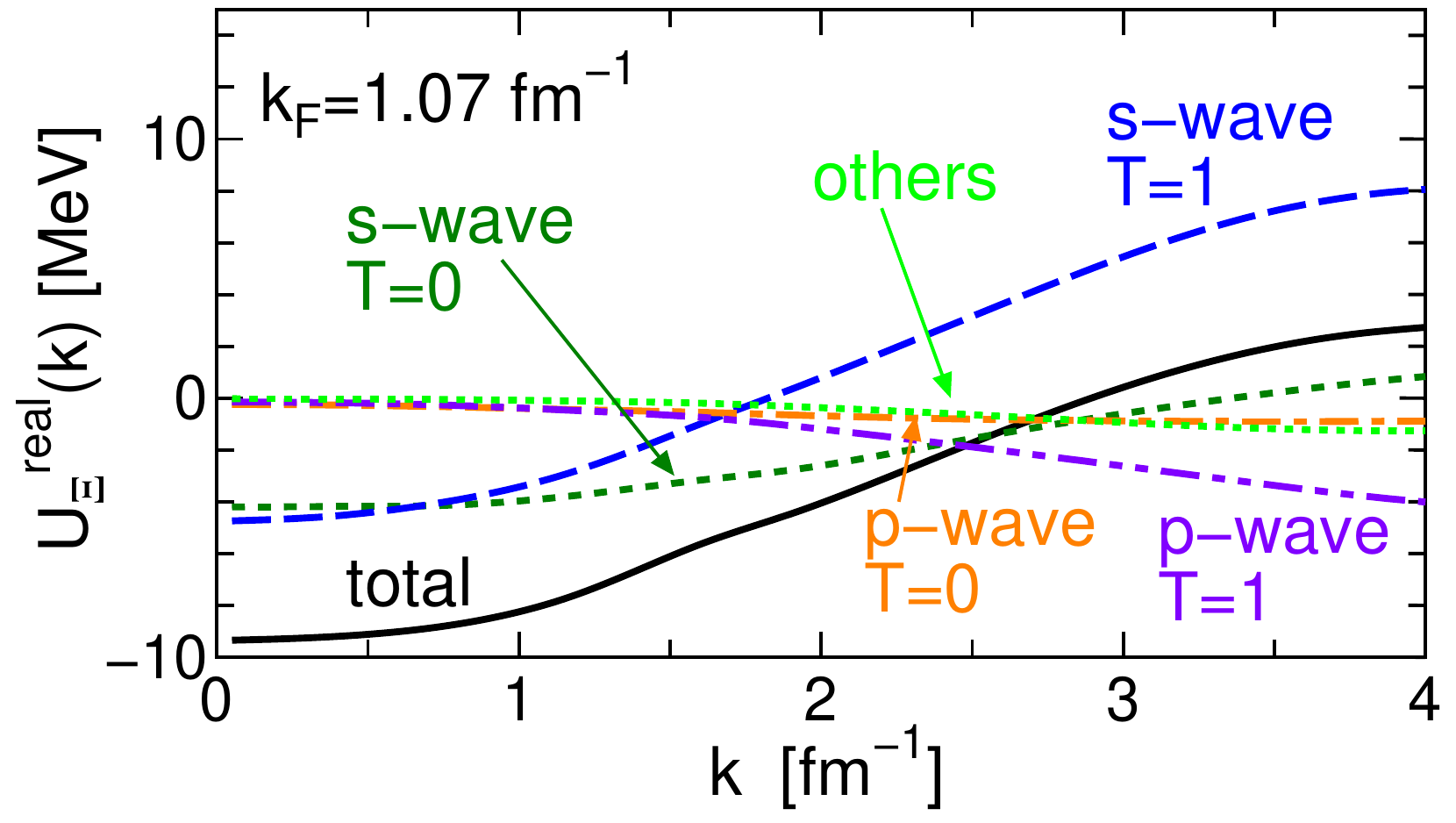}
 \includegraphics[width=0.4\textwidth,clip]{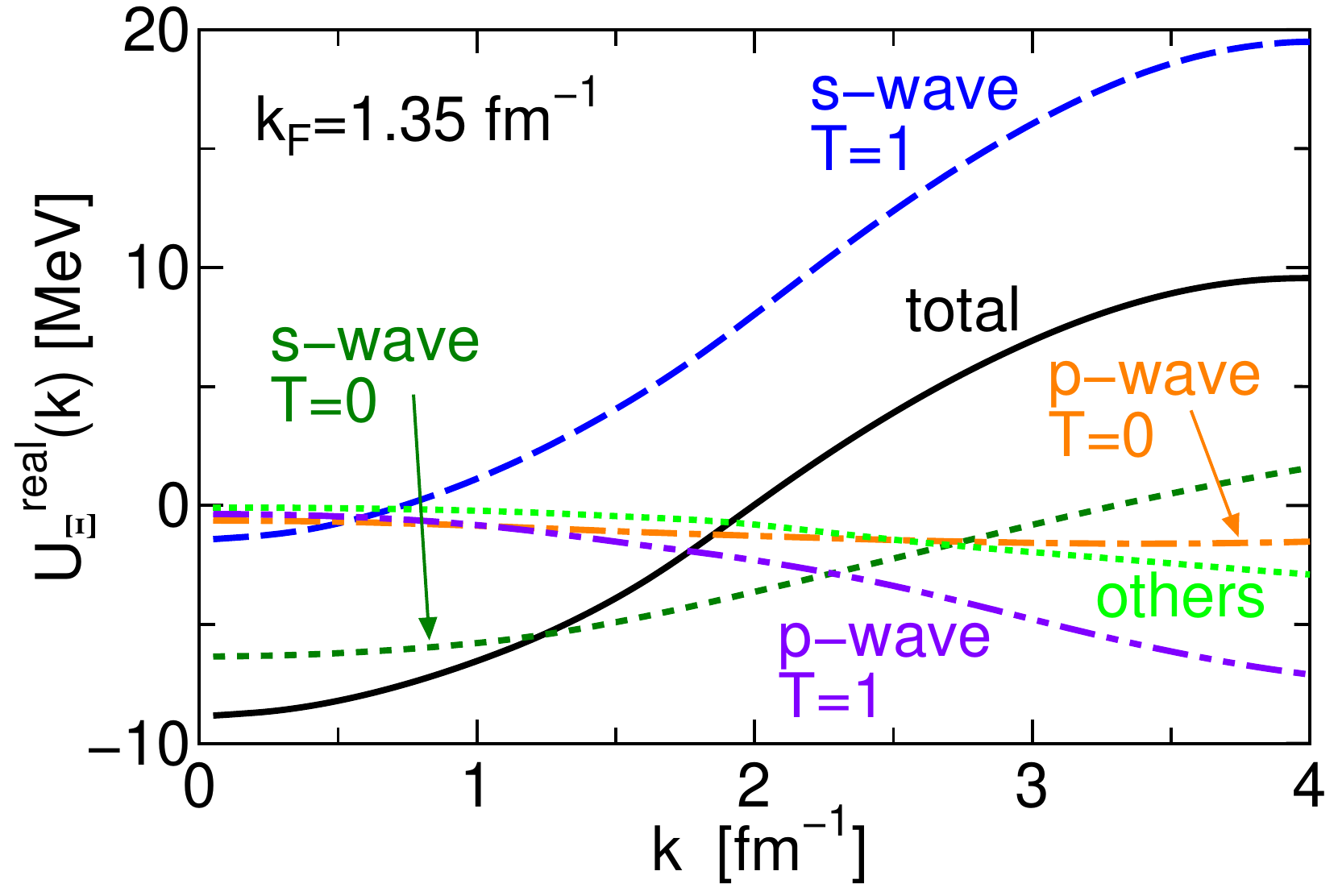}
 \caption{Momentum dependence of the contributions to the $\Xi$ single-particle potential
from the $s$, $p$, and other partial waves after summing over the total $J$ in the isospin
$T=0$ or $T=1$ channel in SNM with $k_F=1.07$ fm$^{-1}$ and $k_F=1.35$ fm$^{-1}$.}
\label{fig:spkfpw}
\end{figure}

The partial-wave contributions after summing over the total $J$ are presented
in Fig. \ref{fig:spkfpw}. At low densities, the $T=1$ $s$-wave contribution is small, and
the $T=0$ $s$-wave contribution predominates to make the $\Xi$ potential weakly attractive
at low momentum. Increasing the density, the $T=1$ $s$-wave contribution changes sign
from attractive to repulsive. The partial-wave decomposition indicates that the
$\Xi^-$ hyperon potential in high-density neutron star matter is repulsive because
the attractive contribution in the isospin $T=0$ $s$-state is absent.

\section{$\Xi$ potentials in finite nuclei}
It is necessary to consider $\Xi$ potentials explicitly in finite nuclei, to confront with
empirical data of the $\Xi^-$ bound states from the emulsion experiments, the $K^+$
spectra of $(K^-,K^+)$ $\Xi^-$ production inclusive reactions on nuclei, and other
experiments in the future. A useful way to deduce a potential in a finite nucleus from
the potential in nuclear matter is a local-density approximation. This method is simple
but has been demonstrated to be reliable in various applications.

In the present case, the procedure to obtain the coordinate space optical model potential
$U_\Xi (r;E)$ for the $\Xi$ hyperon energy $E$ from the potential $U_\Xi (k;k_F)$ in SNM
is the following. First, in SNM, the momentum dependence is converted to the energy
dependence, $U_\Xi (E,k_F)$, by solving the energy-momentum relation
$E=\frac{\hbar^2}{2m_N}k^2+ U_\Xi (k;k_F)$.
Next, the Fermi momentum $k_F$ is replaced by the local Fermi momentum $k_F(r)$ defined by
the local density of a finite nucleus $\rho(r)$ as $k_F(r)=(3\pi^2 \rho(r)/2)^{1/3}$.
The resulting potential $U_\Xi (E;k_F(r))$ is regarded as an energy-dependent optical model
potential $U_\Xi (r;E)$.

One insufficiency in this procedure is that a finite range effect is lost. The prescription to
remedy the drawback has been known to introduce a convolution of the potential with
a Gaussian form factor with an appropriate range $\beta$ of around 1 fm:
\begin{equation}
 (\sqrt{\pi} \beta)^{-3} e^{-|\br-\br'|^2/\beta^2}, \label{eq:gff}
\end{equation}
which was introduced in Ref. \cite{JLM77} as an improved local-density approximation.
It was checked for a nucleon case that this prescription reproduces very well a microscopic
optical model potential calculated with two-body $G$-matrices \cite{Toyo18}.

The $\Xi$ potentials in finite nuclei obtained by the above method, based on the
$\Xi$ potentials in SNM with the chiral NLO baryon-baryon interactions
in the $S=-2$ sector, are shown in the following subsection.

\subsection{Calculated $\Xi$ potentials in finite nuclei}
\begin{figure}[H]
\centering
 \includegraphics[width=0.4\textwidth,clip]{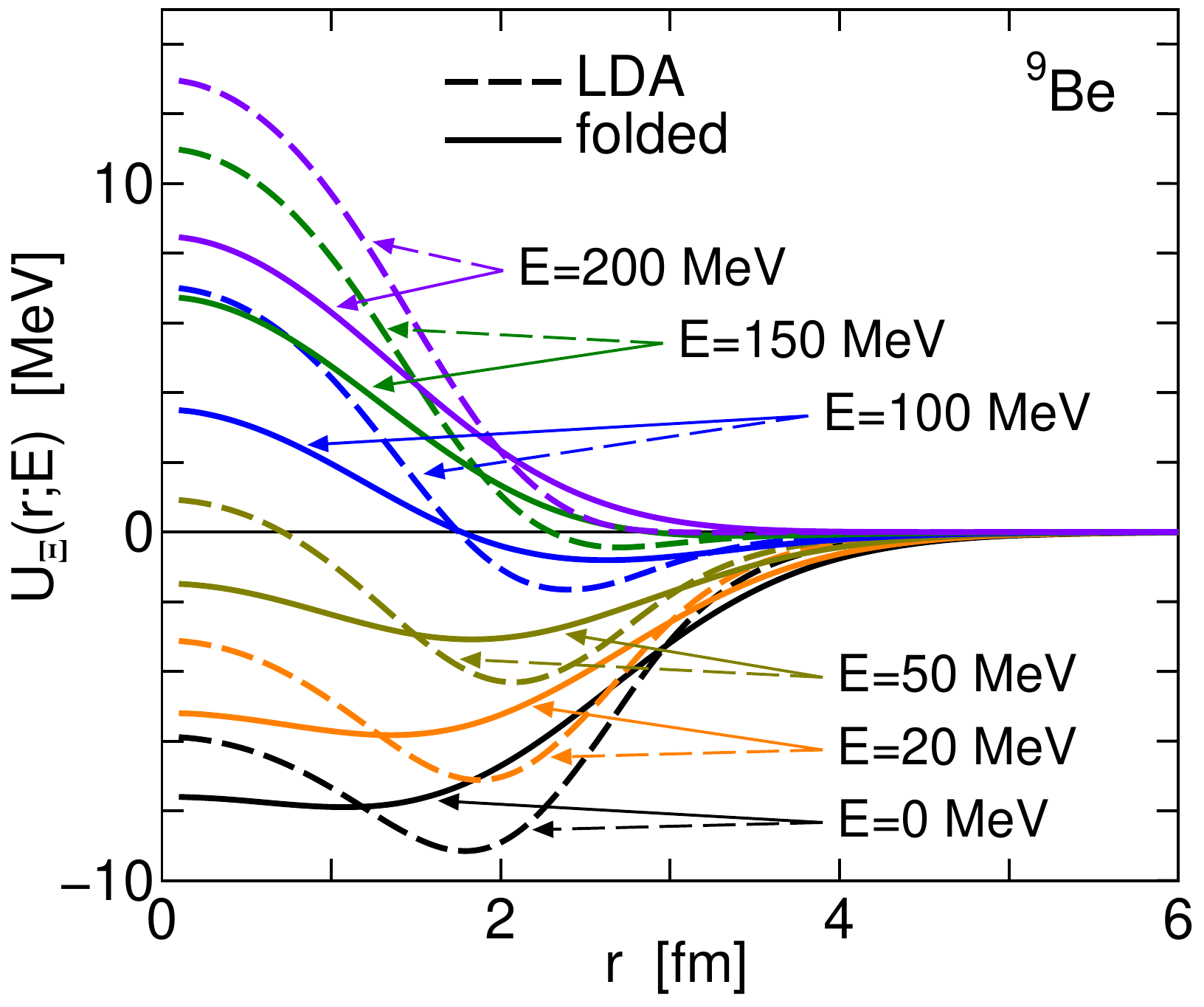}
 \includegraphics[width=0.4\textwidth,clip]{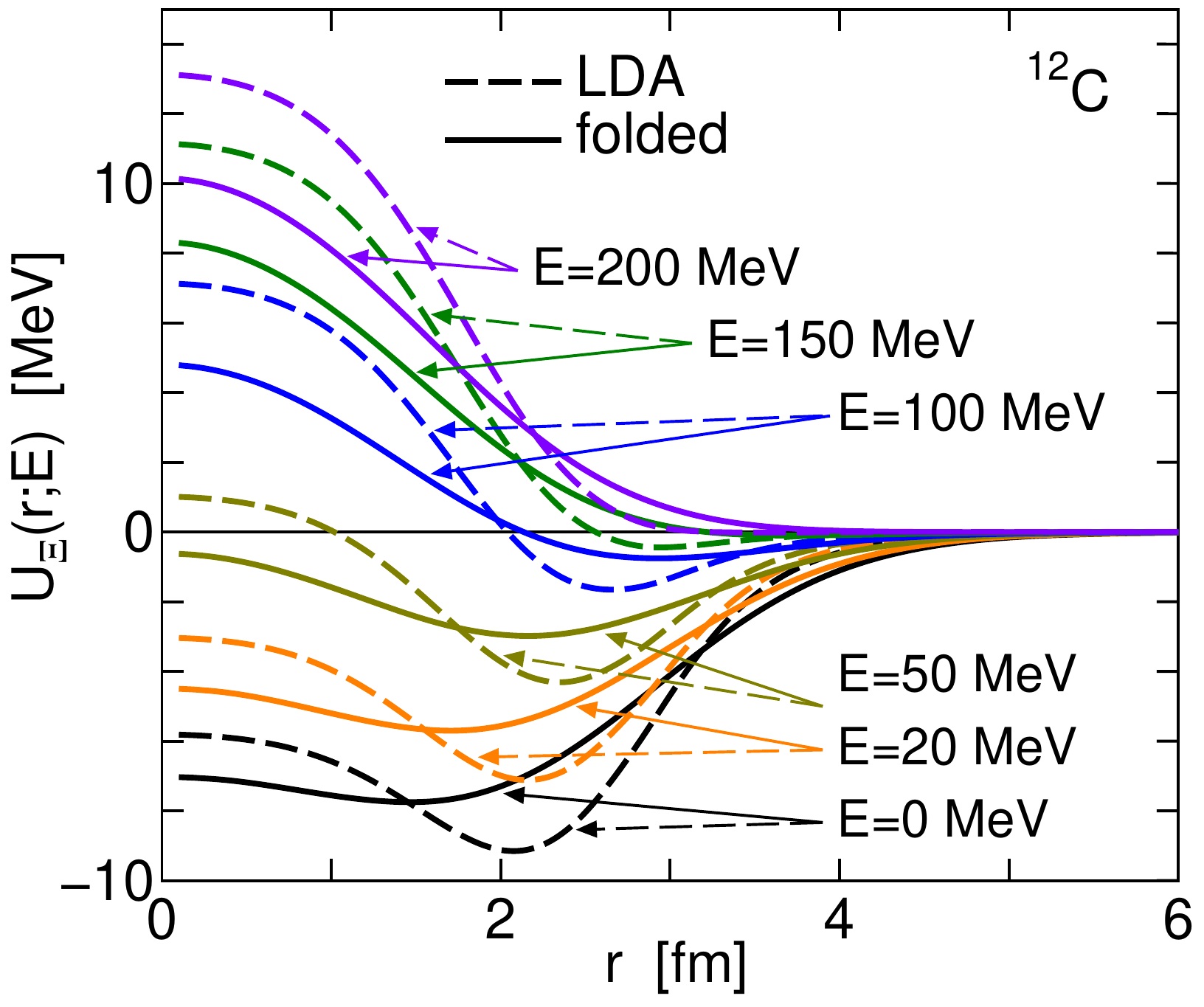}
 \includegraphics[width=0.4\textwidth,clip]{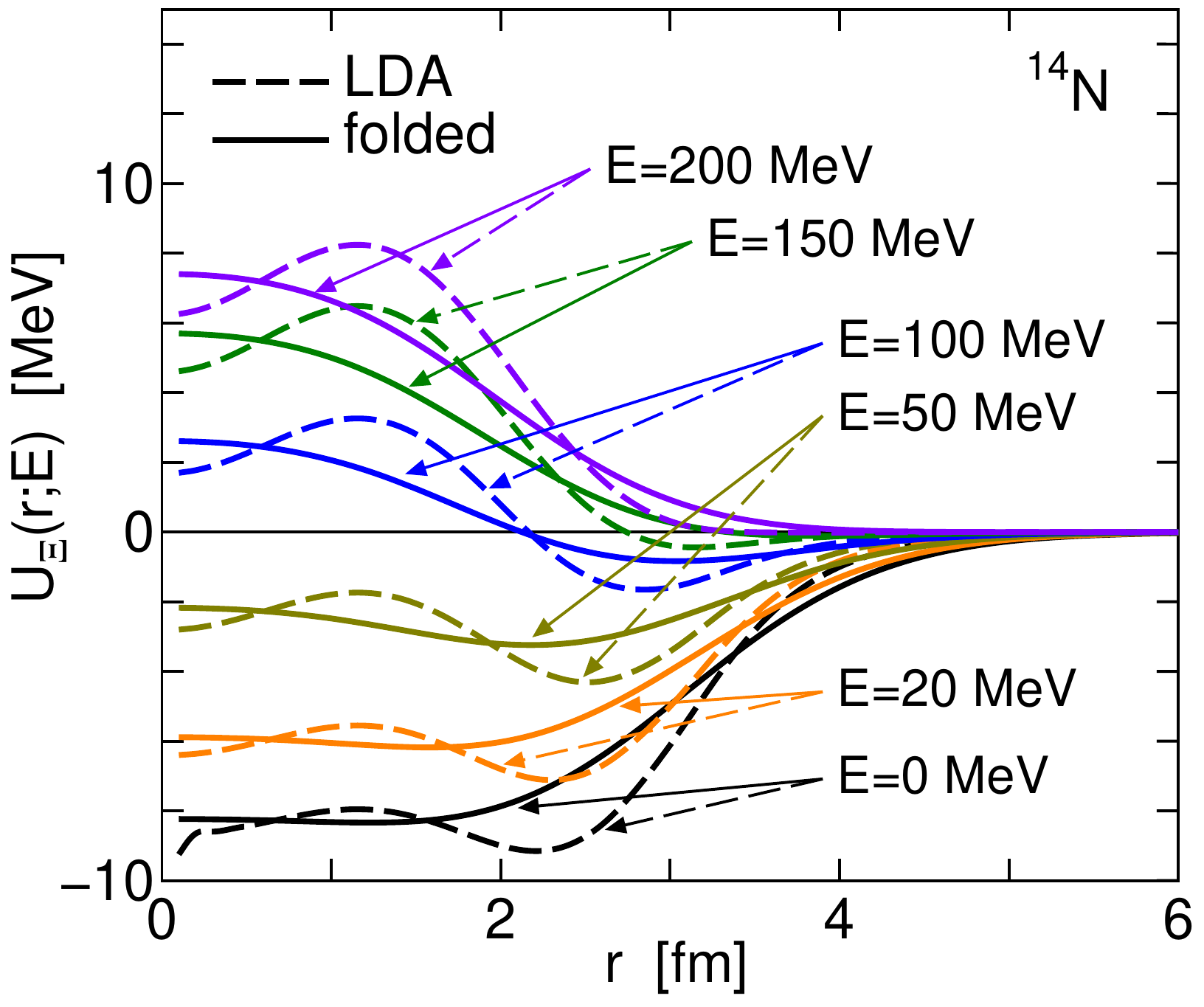}
 \caption{Radial dependence of the $\Xi$ potentials in finite nuclei, $^9$Be, $^{12}$C,
and $^{14}$N, calculated from those in SNM by a local-density approximation for
various $\Xi$ energies $E$. Results before and after folding the Gaussian form
factor of Eq. \ref{eq:gff}
with $\beta=1$ fm are represented by dashed and solid curves, respectively.}
\label{fig:ldasp}
\end{figure}

Calculated coordinate-space $\Xi$ potentials in  $^{9}$Be, $^{12}$C, and $^{14}$N are
shown in Fig. \ref{fig:ldasp} for various $\Xi$ energies from 0 to 200 MeV. The nucleon
density distributions are prepared by density-dependent Hartree-Fock calculations with
the finite-range G-0 force \cite{CS72}. The dashed curves
stand for the results of the local-density approximation before convoluting the Gaussian form
factor of Eq. \ref{eq:gff}. Potentials after applying the Gaussian folding with $\beta=1.0$ fm
are represented by the solid curves. Reflecting the momentum
dependence of the $\Xi$ single-particle potential in SNM, the $\Xi$ potential is attractive
at low-energies and becomes repulsive at higher densities.

It is useful for later applications to parametrize the calculated $\Xi$ potential in a simple
functional form. It is found that the potentials shown in Fig. \ref{fig:ldasp} are simulated
reasonably well by a sum of an attractive part and a repulsive part both in a Woods-Saxon
form. Namely, the calculated $\Xi$ potential is parametrized in the following form with
energy dependent strengths and geometry constants.
\begin{equation}
 V_\Xi (r,E)= \sum_{i=1,2} \frac{V_i}{1+\exp((r-r_i)/a_i)}
\label{eq:ws}
\end{equation}
The fitting is carried out for $^{9}$Be, $^{12}$C, and $^{14}$N, and the resulting
parameters are tabulated in Table I. The energy dependence of geometry parameters
is not large.

Although it should be kept in mind that the $\Xi$ potential calculated by the local-density
approximation bears uncertainties, it is meaningful to investigate the predictions of
the potential for $\Xi$ bound and scattering states.

\begin{table}[thb]
 \caption{Fitted parameters of the $\Xi$ potential in the form of
Eq. \ref{eq:ws} for $^{9}$Be, $^{12}$C, and $^{14}$N. The units of $V_i$, $r_i$, $a_i$ and
the energy $E$ are MeV, fm, fm, and MeV, respectively}
 \setlength{\tabcolsep}{8pt}
 \begin{center}
 \begin{tabular}{l} \hline\hline
  \hspace{10em}$^{9}$Be   \\ \hline
  $V_1=\left\{\begin{array}{ll}-0.0004(E-160.0) \hspace{1em}
   \mbox{for $E\le 160$} \\    0 \hspace{8em}\hspace{1em}
   \mbox{for $E > 160$} \end{array}  \right. $ \\
  $r_1=\left\{\begin{array}{ll}2.62-0.00021E+0.000049E^2 \mbox{for
   $E\le 90$} \\ 5.24-0.058E+0.00037E^2 \hspace{1em}
   \mbox{for $E > 90$} \end{array} \right.$ \\
  $a_1=\min(0.59-0.0005E, 1.26-0.0074E)$ \\
  $V_2=\left\{\begin{array}{ll}3.71-0.042E+0.0009E^2\hspace{1em}
   \mbox{for $E < 40$} \\ 2.15+0.033E \hspace{6em}
   \mbox{for $E \ge 40$} \end{array}  \right. $ \\
  $r_2=\min(0.875+0.86E, 1.06+0.002E)$ \\
  $a_2=\left\{\begin{array}{ll}0.96-0.22E+0.00033E^2 \mbox{ for $E\le 20.5$} \\
  0.72-0.0038E+0.000013E^2 \mbox{ for $E > 20.5$} \end{array} \right.$ \\\hline
  \hspace{10em}$^{12}$C  \\ \hline
  $V_1=\left\{\begin{array}{ll}-0.0003(E-190.0) \hspace{1em}
   \mbox{ for $E\le 190$} \\    0 \hspace{8em}\hspace{1em}
   \mbox{ for $E > 190$} \end{array}  \right. $ \\
  $r_1=\left\{\begin{array}{ll} \min(2.86-0.0008, 2.45+0.008E) \mbox{ for $E\le 110$} \\
     4.61-0.031E+0.000177E^2  \mbox{ for $E > 110$} \end{array} \right.$ \\
  $a_1=0.575+0.00084E-0.000017E^2$ \\
  $V_2=4.50+0.032E$ \\
  $r_2=\min(1.38+0.003E, 1.53)$ \\
  $a_2=\max(0.864-0.0059E,0.696-0.00095E,$ \\
   \hspace{5em}$0.436+0.00049E)$ \\\hline
  \hspace{10em}$^{14}$N  \\ \hline
  $V_1=\left\{\begin{array}{ll}-0.00025\times(E-190.0) \hspace{1em}
   \mbox{for $E\le 190$} \\    0 \hspace{11em}
   \mbox{for $E > 190$} \end{array}  \right. $ \\
  $r_1= 3.1+0.0035E$ \\
  $a_1=0.55+0.00038$ \\
  $V_2=\min( 0.75+0.046E,2.21+0.027E)$ \\
  $r_2=\min(1.23+0.0091E, 1.71+0.0013E)$ \\
  $a_2=\min(0.38+0.014E, 0.76-0.0016E)$ \\\hline
 \end{tabular}
 \end{center}
\label{table:para}
\end{table}

\subsection{$\Xi$ bound states in $^{12}$C and  $^{14}$N}
The attractive $\Xi$ potential at low energies shown in Fig. \ref{fig:sppt} indicates that
$\Xi^-$ atomic bound states are lowered. In this subsection, single-particle energies of
the $\Xi$ potentials obtained with the NLO ChEFT interactions are discussed for
$^{12}$C and $^{14}$N. Because the imaginary part of the $\Xi$ potential is very small
at low energies as seen in Fig. \ref{fig:sppt}, the imaginary part is disregarded.
The Coulomb attraction for the $\Xi^-$ hyperon is taken into account by the
potential of uniform charge distribution with a radius of $R_c=1.15A^{1/3}$ fm:
\begin{equation}
 V_c(r)=\left\{ \begin{array}{cc} -\frac{Ze^2}{2R_c} \left[ \frac{3}{2}-\frac{r^2}{2R_c^2} \right]
 & \mbox{ for }r<R_c, \\
   -\frac{Ze^2}{r} & \mbox{ for }r \ge R_c,
\end{array} \right.
\end{equation}
where $A$ and $Z$ are the mass number and the proton number of the target
nucleus, respectively, and $e$ is an elementary charge. In the following evaluations,
the $\Xi^-$ and $\Xi^0$ masses are set to be 1321.3 and 1314.8 MeV, respectively,
though the average value of 1318.1 MeV is used in SNM calculations.

Experimentally, possible $\Xi^-$-$^{12}$C bound states were suggested \cite{Aok09}
by the events in the nuclear emulsion experiments at KEK. The diagnosed energies are
$3.89\pm0.24$, $2.84\pm0.15$, and $0.82\pm0.17$ MeV. Then, the first evidence of
a deeply bound $\Xi^-$ state in $^{14}$N was reported in Ref. \cite{Naka15} from the
emulsion data at KEK-PS. The event was identified as the process
$\Xi^- + \mbox{$^{14}$N} \rightarrow \mbox{$^{10}_\Lambda$Be}+ \mbox{$^5_\Lambda$He}$
and the binding energy of the $\Xi^-$ state was assigned to be $4.38\pm0.25$ or
$1.11\pm0.25$ MeV, depending on the choice of the excitation energy of $^{10}_\Lambda$Be.

Calculated $\Xi$ single-particle energies and radii for $^{12}$C and $^{14}$N with
the potential $U_{\Xi}(r;E)$ numerically evaluated at $E=-5$ MeV and $E=0$ MeV are
given in Table \ref{table:lda}. The difference between the single-particle energies at  $E=-5$
MeV and $E=0$ MeV indicates the energy dependence of the $\Xi$ potential. It is
possible for the NLO ChEFT $\Xi$ potential to generate a $0s$ $\Xi^0$ bound state
in $^{12}$C and $^{14}$N, but no $0p$ bound state exists. For $\Xi^-$, the $0s$ state
appears, with the assistance of the Coulomb attraction,  at $\approx -5$ MeV in $^{14}$N and
$\approx -4$ MeV in $^{12}$C. The $0p$ state is slightly lowered from the atomic state and
the $0d$ state is hardly affected by the $\Xi$-nucleus potential.
It is interesting to observe that the deeply bound $\Xi^-$ state of the $\Xi^-$-$^{14}$N
system at $4.38\pm 0.25$ MeV found in Ref. \cite{Naka15} nearly matches
the $0s$ state in the present calculation. Even another assignment of the $\Xi^-$ binding
energy of $1.11\pm 0.25$ MeV roughly corresponds to the $0p$ state in Table II.
The experimental candidates of the $\Xi^-$ bound state in $^{12}$C at 3.89, 2.84 or 0.82 MeV
\cite{Aok09} are not far from the calculated $0s$ and $0p$ states given in Table \ref{table:lda}.

\begin{table}[htb]
 \caption{Energies and radii of $\Xi$ bound states yielded by the potential $U_\Xi(r;E)$ numerically
obtained through the Gaussian-folded LDA at $E=-5$ and $0$ MeV. The entry values of the energy
and the root-mean-square radius $\sqrt{\langle r^2\rangle}$ are in MeV and fm, respectively. }
\setlength{\tabcolsep}{7pt}
 \begin{center}
 \begin{tabular}{cccccc} \hline\hline
 &  & \multicolumn{2}{c}{$U_\Xi (r,E=-5)$} & \multicolumn{2}{c}{$U_\Xi (r,E=0)$} \\
 nucleus & state & energy & $\sqrt{\langle r^2\rangle}$  & energy &
 $\sqrt{\langle r^2\rangle}$ \\ \hline
 $^{12}$C & $\Xi^0$ 0$s$ &  $-1.18$ & $\phantom{0}4.12$  & $-1.01$ & $\phantom{0}4.34$ \\
 & $\Xi^-$ 0$s$ &  $-4.22$ & $\phantom{0}3.20$  & $-3.99$ & $\phantom{0}3.29$ \\
 & $\Xi^-$ 0$p$ &  $-0.35$ & $16.00$ & $-0.34$ & $16.11$ \\
  & $\Xi^-$ 0$d$ &  $-0.13$ & $42.74$ & $-0.13$ & $42.74$ \\ \hline
 $^{14}$N & $\Xi^0$ 0$s$ &  $-1.77$ & $\phantom{0}3.70$  & $-1.56$ & $\phantom{0}3.84$ \\
 & $\Xi^-$ 0$s$ &  $-5.40$ & $\phantom{0}3.03$ & $-5.14$ & $\phantom{0}3.10$ \\
  & $\Xi^-$ 0$p$ &  $-0.67$ & $9.04$ & $-0.64$ & $\phantom{0}9.62$ \\
  & $\Xi^-$ 0$d$ &  $-0.17$ & $36.10$ & $-0.17$ & $36.10$\\ \hline \hline
 \end{tabular}
 \end{center}
\label{table:lda}
\end{table}

\begin{figure}[bht]
\centering
 \includegraphics[width=0.4\textwidth,clip]{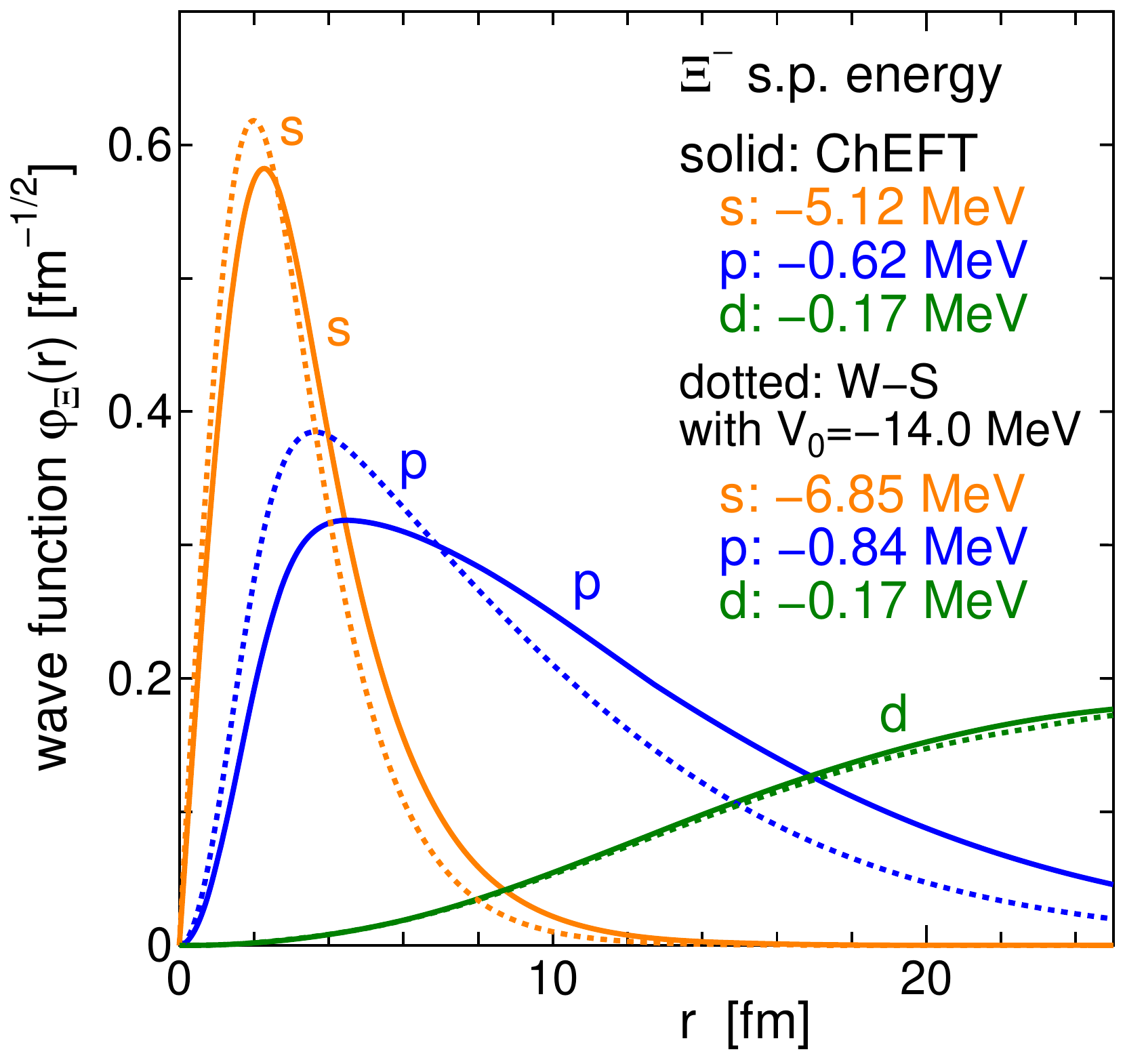}
 \caption{$\Xi$ wave functions of the $0s$, $0p$, and $0d$ bound states given
in Table \ref{table:ws} for the $\Xi^-$-$^{14}$N system.
Solid curves represent results of the potential given in Eq. (\ref{eq:ws}) with $E=0$.
For comparison, results with a single Woods-Saxon potential with the depth
of $-14$ MeV and the geometry of $r_0=1.1 A^{1/3}$ and $a_0=0.65$ are shown
by dotted curves. The dotted curve for the $0d$ state is almost indistinguishable
from the solid curve.
}
\label{fig:wf}
\end{figure}

It is informative to present $\Xi^-$ single-particle energies and radii of the $0s$, $0p$,
and $0d$ $\Xi^-$ bound states in $^{12}$C and $^{14}$N for the potential given by the
parameters at $E=0$ in Table \ref{table:para}. They are given in Table \ref{table:ws}.
The results of $V_\Xi (r,E=0)$ are close to those in Table \ref{table:lda}. Their wave
functions in $^{14}$N are shown in Fig. \ref{fig:wf}.The bound states expected by
a Woods-Saxon potential with the canonical depth of 14 MeV and the
geometry parameters of $r_0=1.1A^{1/3}$ fm and $a=0.65$ fm are also included
in Table \ref{table:ws} for comparison. The corresponding wave functions are also
included in Fig. \ref{fig:wf}. Unless a deeper bound state than $4.38\pm0.25$ MeV in $^{14}$N
or $3.89\pm 0.24$ MeV is found in future experiments, the depth of 14 MeV
in a standard Woods-Saxon form is not probable for the $\Xi$-nucleus potential.

Possible $\Xi^-$ bound states as well as atomic energy level shifts in heavier nuclei
such as $^{40}$Ca and beyond are interesting to discuss. However, this subject is left for
future studies because it was difficult to parametrize the potential calculated by the
local-density approximation in the easy-to-use form of Eq. \ref{eq:ws} because of
a non-monotonic shape of the nucleon density distribution of those nuclei.

\begin{table}[bht]
 \caption{Energies and radii of $\Xi$ bound states yielded by the potential $V_{\Xi}(r,E)$ given
by Eq. (\ref{eq:ws}) with the parameters at $E=0$ in Table I. For comparison, those obtained
by a single Woods-Saxon potential with the depth of 14 MeV, the range of $r_1=1.1A^{1/3}$
fm and the surface thickness of $a_1=0.65$ fm are given. The entry values of the energy
and the root-mean-square radius $\sqrt{\langle r^2\rangle}$ are in MeV and fm, respectively. }
\setlength{\tabcolsep}{7pt}
 \begin{center}
 \begin{tabular}{cccccc} \hline\hline
 &  & \multicolumn{2}{c}{$V_\Xi (r,E=0)$} & \multicolumn{2}{c}{W-S} \\
 nucleus & state & energy & $\sqrt{\langle r^2\rangle}$  & energy &
 $\sqrt{\langle r^2\rangle}$ \\ \hline
 & 0$s$ &  $-4.07$ & $\phantom{0}3.27$  & $-5.83$ & $\phantom{0}2.79$ \\
 $^{12}$C & 0$p$ &  $-0.35$ & $16.00$ & $-0.43$ & $12.36$ \\
  & 0$d$ &  $-0.13$ & $41.77$ & $-0.13$ & $42.61$ \\ \hline
 & 0$s$ &  $-5.11$ & $\phantom{0}3.10$ & $-6.85$ & $\phantom{0}2.71$ \\
 $^{14}$N & 0$p$ &  $-0.62$ & $9.84$ & $-0.83$ & $\phantom{0}7.51$ \\
  & 0$d$ &  $-0.17$ & $36.13$ & $-0.17$ & $35.93$\\ \hline \hline
 \end{tabular}
 \end{center}
\label{table:ws}
\end{table}

\begin{figure}[bht]
\centering
 \includegraphics[width=0.4\textwidth,clip]{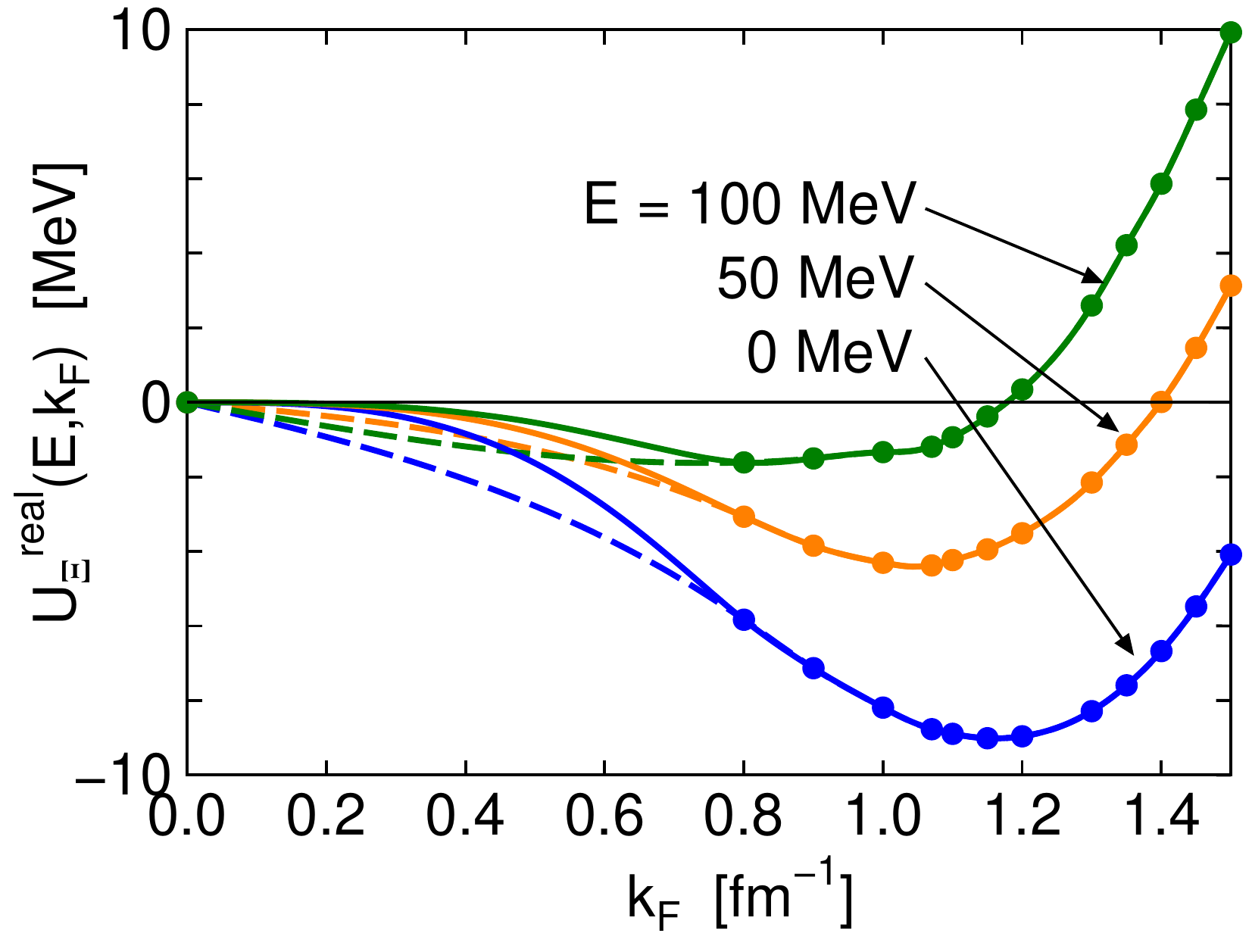}
 \caption{$\Xi$ potential values $U_\Xi(E,k_F)$ calculated in SNM at various
Fermi momenta from $k_F=0.8$ to $1.5$ fm$^{-1}$ are shown by the dots
for the $\Xi$ energies of $E=0$, 50, and 100 MeV. The solid curves represent
the cubic spline fit as a function of the density $\rho=\frac{2k_F^3}{3\pi^2}$.
The dashed curves show the results of the cubic spline fit as a function of
the Fermi momentum $k_F$. The interpolation as a function of $k_F$ tends
to yield a more attractive potential at low densities than that with respect
to the density $\rho$.     
}
\label{fig:dkffit}
\end{figure}

The following remarks about the uncertainties in an interpolation prescription
may be instructive. G-matrix calculations in SNM below the
Fermi momentum $k_F\simeq 0.8$ fm$^{-1}$ are not reliable already
in the nucleon-nucleon case because stable self-consistency of the single-particle
potential is not achieved. The reason is probably that homogeneous nuclear matter
is unstable at low densities due to some cluster formation. In fact, the binding energy
per nucleon in SNM is $7-8$ MeV at $k_F=0.8$ fm$^{-1}$, which is comparable to
that of the $\alpha$ particle. The low-density behavior of the
potential $U_\Xi (E;k_F)$, which is needed in the local-density approximation in a
surface area, relies on interpolation with the condition $U_\Xi(E;k_F=0)=0$.
Thus, the results depend on the interpolation method. In the present evaluation,
$U_\Xi(k;k_F)$ is interpolated by a cubic spline prescription as a function of the density
$\rho=\frac{2k_F^3}{3\pi^2}$ because a standard $t\rho$ prescription is expected to
hold with $t$ being a scattering amplitude.
The results of the fitting are shown in Fig. \ref{fig:dkffit} by solid curves for
$E=0$, $50$, and $100$ MeV. However, if the interpolation is made
as a function of the Fermi momentum $k_F$, more attractive
potentials at low-densities are expected for $E=0$ as shown by dashed curves.
If the latter interpolation method is employed, the $\Xi$ potential becomes
more attractive at low densities, and the $0s$ and $0p$ $\Xi^-$ bound-states appear
at lower energies than those in Table II. The $\Xi$ potential at low densities, which is
responsible for the shift of $\Xi^-$ atomic levels and which can be experimentally
measured, deserves further theoretical investigations.
More experimental data in the future are naturally needed to confirm the reality
of the $\Xi$ potential in finite nuclei.

\section{Inclusive $(K^-,K^+)$ $\Xi$ production spectrum}
Possible $\Xi$ bound states in nuclei have been sought via $(K^-, K^+)$
experiments. The inclusive $K^+$ spectrum of $\Xi^-$ production
around the threshold on $^{12}$C was measured at KEK \cite{Fuk98} and
BNL \cite{Kha00}. No peak structure was observed in these spectra.
Nevertheless, an attractive $\Xi$-nucleus potential in a standard
Woods-Saxon form with a depth of about 14 MeV has been inferred \cite{Kha00}.
However, there was another independent analysis \cite{KH10} which indicated
that the null $\Xi$ potential may be consistent with the BNL data.

The quasi-free peak position and the shape of the $K^+$ spectrum can generally
provide basic information about the $\Xi$-nucleus interaction. Tamagawa
presented, in his thesis \cite{Tam00}, double-differential cross section data of
the quasifree $\Xi^-$ production in the $(K^-, K^+)$ reaction on $^9$Be, which has
not been inspected theoretically. New $(K^-, K^+)$ experiments with an improved
resolution were carried out at KEK \cite{Nag18} and the data are now under analysis.
The whole $K^+$ spectrum from the $\Xi^-$ threshold to beyond the quasifree peak
in the experiments should offer valuable information on the $\Xi$-nucleus
potential as well as $\Xi^-$ bound states.
   
In this section, $K^+$ spectra in $(K^-, K^+)$ $\Xi^-$ production inclusive reactions
on $^9$Be and $^{12}$C are calculated, using the $\Xi^-$-nucleus potential in a
parametrized form obtained in the previous section. The calculations are carried
out by employing the semiclassical distorted wave (SCDW) method \cite{Hash08}.
The outline of the SCDW method is presented in the following. More details
are found in Refs. \cite{Hash08,KH10}.

The basic expression of the distorted wave impulse approximation for the $K^+$ double
differential cross section of the $(K^-,K^+)$ $\Xi^-$ production inclusive reaction
is given by
\begin{align}
\frac{d^2 \sigma}{dW d\Omega} &=\frac{\omega_{i,red}\omega_{f,red}}{(2\pi )^2}
 \frac{p_f}{p_i} \sum_{p,h}\; \delta (W-\epsilon_p + \epsilon_h ) \notag \\
 & \times\frac{1}{4\omega_i \omega_f} | \langle \chi_f^{(-)} \phi_p^{(-)} |v_{f,p,i,h}|
 \chi_i^{(+)} \phi_h \rangle |^2.
\end{align}
The semiclassical approximation for the wave functions of the incoming and outgoing
kaons means
\begin{align}
 \chi_{i,f}^{(\pm)}(\br) \chi_{i,f}^{(\pm)*}(\br') 
 &= \chi_{i,f}^{(\pm)}\left(\bR+\frac{1}{2}\bs\right) \chi_{i,f}^{(\pm)*}
 \left(\bR-\frac{1}{2}\bs\right) \notag \\
 &\simeq |\chi_{i,f}^{(\pm)}(\bR)|^2 e^{i\bk_{\i,f}(\bR)\cdot\bs},
\end{align}
where the direction and the magnitude of the local momentum $\bk_{i,f}(\bR)$ are
determined by the momentum density $\bk_{i,f}^{\pm}$ and the local energy-momentum
relation, respectively:
\begin{align}
 \bk_{i,f}^{\pm} =& \frac{\Re \{\chi_{i,f}^{\pm *}(\bR)(-i)\nabla\chi_{i,f}^{\pm}(\bR)\}}
 {|\chi_{i,f}^{\pm}(\bR)|^2},
\end{align}
and
\begin{align}
 m_K^2 +\bk_{i,f}^2 (\bR) + 2\omega_{i,f}[U^{real}(\bR) +V_{Coul}(\bR)] \notag \\
   -V_{Coul}^2(R) =\omega_{i,f}^2.
\end{align}
The introduction of the above approximation together with the similar approximation
for the $\Xi^-$ wave function and the Wigner transformation $\Phi_h$ of the nucleon
single-particle wave function of the target nucleus lead to the following expression:
\begin{align}
\frac{d^2 \sigma}{dW d\Omega} & = \frac{\omega_{i,red} \omega_{f,red}}{(2\pi )^2}
 \frac{p_f}{p_i} \xi^6 \int \int d\bR d\bK \sum_{p}\;
 \frac{1}{4\omega_i \omega_f} \notag \\
 &\times |\chi_f^{(-)}(\bR)|^2 |\chi_i^{(+)}(\bR)|^2  |\phi_p^{(-)}(\xi \bR)|^2
  |v_{f,p,i,h}|^2\notag \\
 & \times  \frac{(2\pi)^3}{\xi^3}  \sum_h \Phi_h\left(\xi\bR,\frac{1}{\xi}\bK\right)
  \delta (W-\epsilon_p + \epsilon_h ) \notag \\
 & \times \delta(\bK +\bk_i(\bR)-\bk_f(\bR) -\bk_p(\bR)).
\end{align}
In this method, a factorization approximation is not introduced for the elementary
amplitude, and the energy angle dependences are treated explicitly, though
the on-shell approximation has to be admitted.

The distorted waves $\chi_{i,f}^{\pm}$ of the incoming $K^-$
and outgoing $K^+$ are described by the Klein-Gordon equation with the kaon-nucleus
potential in a $t\rho$ prescription.  The explicit construction of the kaon-nucleus
potential and the parametrization of the $K^- p\rightarrow K^+\Xi^-$ elementary
amplitude are described in Ref. \cite{Hash08}. The introduction of the Wigner
transformation $\Phi_h$ takes into account the Fermi motion of the nucleon
in applying the $K^- p\rightarrow K^+\Xi^-$ elementary amplitude. It is noted that
the wave functions for constructing $K^-$ and $\Xi^-$ potentials are assumed
to be the same as those of the target in the calculations presented below,
although the $(K^-,K^+)$ reaction changes the target to a different nuclide.

\subsection{$^9$Be$(K^-,K^+)$ spectrum}
\begin{figure}[H]
\centering
 \includegraphics[width=0.4\textwidth,clip]{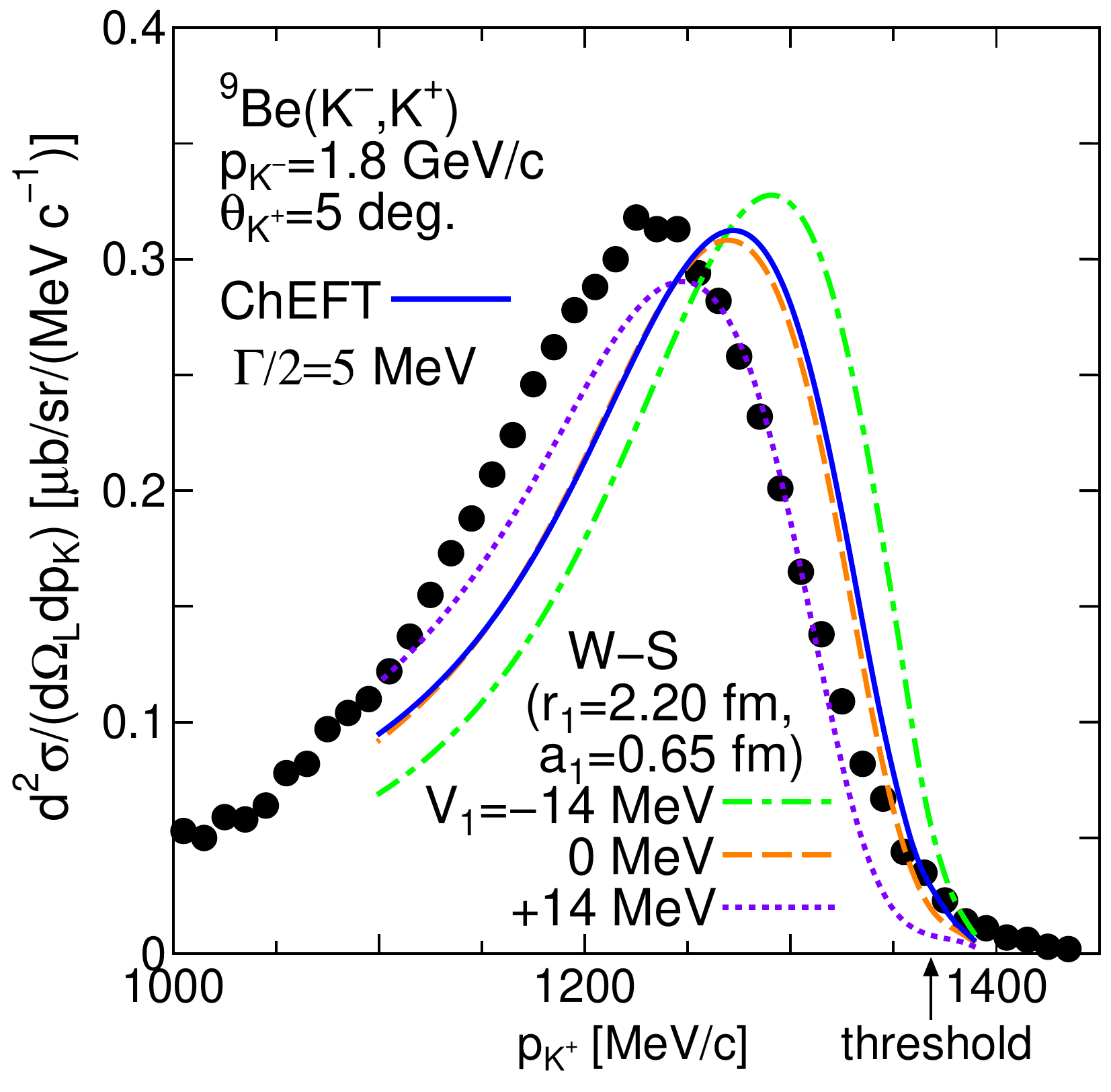}
 \caption{$K^+$ spectrum of $(K^-,K^+)$ $\Xi$ inclusive production
reaction on $^{9}$Be. The incident $K^-$ momentum is $p_{K^-}=1.8$ GeV/c.
The reaction angle in the laboratory frame is set to be $\theta_{K^+}=5^\circ$.
The width of $\Gamma/2=5$ MeV is applied for Lorentz-type smearing.
The experimental resolution of $\Delta E=6.1$ MeV is also taken into account,
as explained in the text. Experimental data are taken from Ref. \cite{Tam00},
which is an average over $1.5^\circ < \theta_{K^+} <8.5^\circ$.}
\label{fig:be9}
\end{figure}

Calculated $K^+$ spectra of $(K^-,K^+)$ $\Xi^-$ production inclusive reactions
on $^9$Be for the incident $K^-$ momentum of 1.8 GeV/c are shown
in Fig. \ref{fig:be9}, compared with the experimental data provided
by Tamagawa \cite{Tam00}. Because the data was an average over the $K^+$
angle between $1.5^\circ$ and $8.5^\circ$, calculations are carried out
at $\theta_{K^+}=5^\circ$. Besides the ChEFT potential, results are presented
with three cases of a single Woods-Saxon potential, the strengths of $V_1 =-14,0$,
and $+14$ MeV with the geometry parameters of $r_1=2.20$ fm, and $a_1=0.65$ fm.
To simulate the effect of the imaginary potential, Lorentz-type smearing with
a width of $\Gamma/2=5$ MeV is introduced. The width should be energy dependent,
but the typical value of 5 MeV is chosen to discuss the wide energy range of
the spectrum. More elaborate treatment of the width in the SCDW
method is a future problem. In addition, the experimental resolution is taken into
account by convoluting the spectrum with a Gaussian function:
\begin{equation}
 f(E,\Delta E)=\frac{1}{\Delta E} \sqrt{\frac{\log 2}{\pi}} e^{-\log 2 (E/\Delta E)^2}.
\end{equation}
The resolution $\Delta E$ is set to be 6.1 MeV, corresponding to the experimental
momentum resolution of $(\Delta p/p)_{K^+}=1$ \% \cite{Tam00}.
 
First, it is emphasized that cross sections are almost reproduced in their magnitude
without any multiplicative factor. The position of the quasi-free peak reveals
important information about the feature of the $\Xi$ potential. It is seen in
Fig. \ref{fig:be9} that the peak position systematically shifts to the smaller
$p_{K^+}$, namely to the higher $\Xi^-$ energy, with changing the
Woods-Saxon strength from attractive to repulsive. The $\Xi^-$ potential with
the NLO ChEFT potential is attractive at low energies and becomes repulsive
as the energy increases. This energy-dependent potential provides a spectrum
close to that of the null potential. Observing that the potential
with $V_1=14$ MeV works rather well in the higher $\Xi^-$ energy region, the ChEFT
potential may need more repulsive character.

It is also to be kept in mind that, because the shape of the response function is affected
by the momentum dependence of the $K^\pm$ and $\Xi$ potentials, nonlocality
effects \cite{Koh83} for these hadrons are worthwhile to investigate in the future
analysis.

\subsection{$^{12}$C$(K^-,K^+)$ spectrum}
\begin{figure}[tbh]
\centering
 \includegraphics[width=0.4\textwidth,clip]{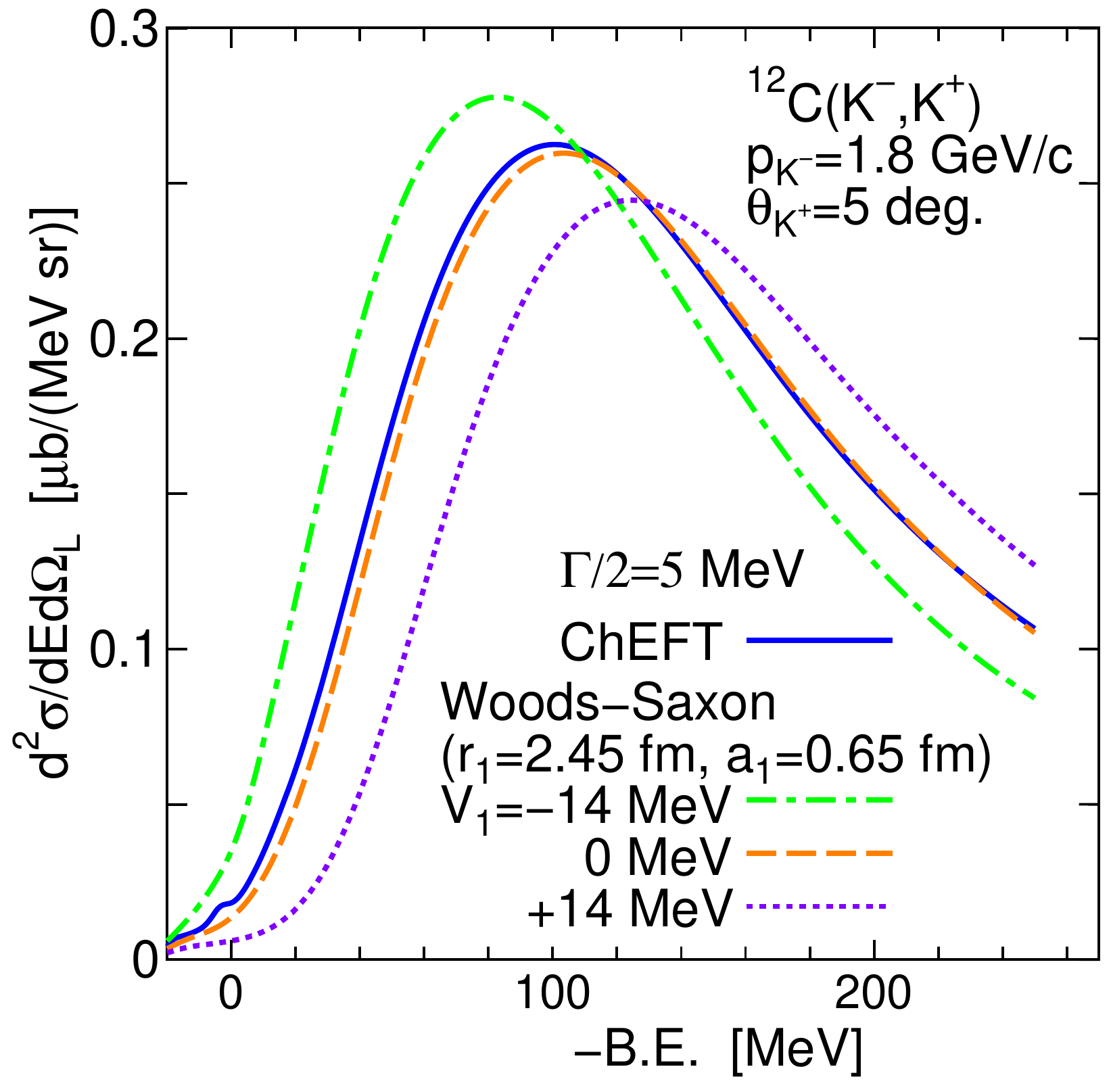}
 \caption{$K^+$ spectrum of $(K^-,K^+)$ $\Xi$ inclusive production
reactions on $^{12}$C. The incident $K^-$ momentum is $p_{K^-}=1.8$ GeV/c and
the reaction angle in the laboratory frame is $\theta_{K^+}=5^\circ$.
The width of $\Gamma/2=5$ MeV is applied for Lorentz-type
smearing. The experimental resolution of $\Delta E-2$ MeV is assumed.}
\label{fig:c12}
\end{figure}

Calculated $K^+$ spectra of $(K^-,K^+)$ $\Xi^-$ production inclusive reactions on $^{12}$C
for the incident $K^-$ momentum of 1.8 GeV/c with $\theta_{K^+}=5^{\circ}$ are shown
in Fig. \ref{fig:c12}. Results employing single Woods-Saxon potentials of $V_1=-14$, 0,
and $+14$ MeV with $r_1=2.45$ and $a_1=0.65$ as well as the ChEFT potential given
by Eq. (\ref{eq:ws}) are presented. Here, the effects of the imaginary potential are
simulated by convoluting the spectrum with a Lorentz-type distribution function.
In Fig. \ref{fig:c12}, the half-width is set to be $\Gamma/2=5$ MeV. Considering the
resolution of the KEK experiments \cite{Nag18}, the Gaussian smearing by Eq. (9)
with $\Delta E=2$ MeV is introduced.

\begin{figure}[tbh]
\centering
 \includegraphics[width=0.4\textwidth,clip]{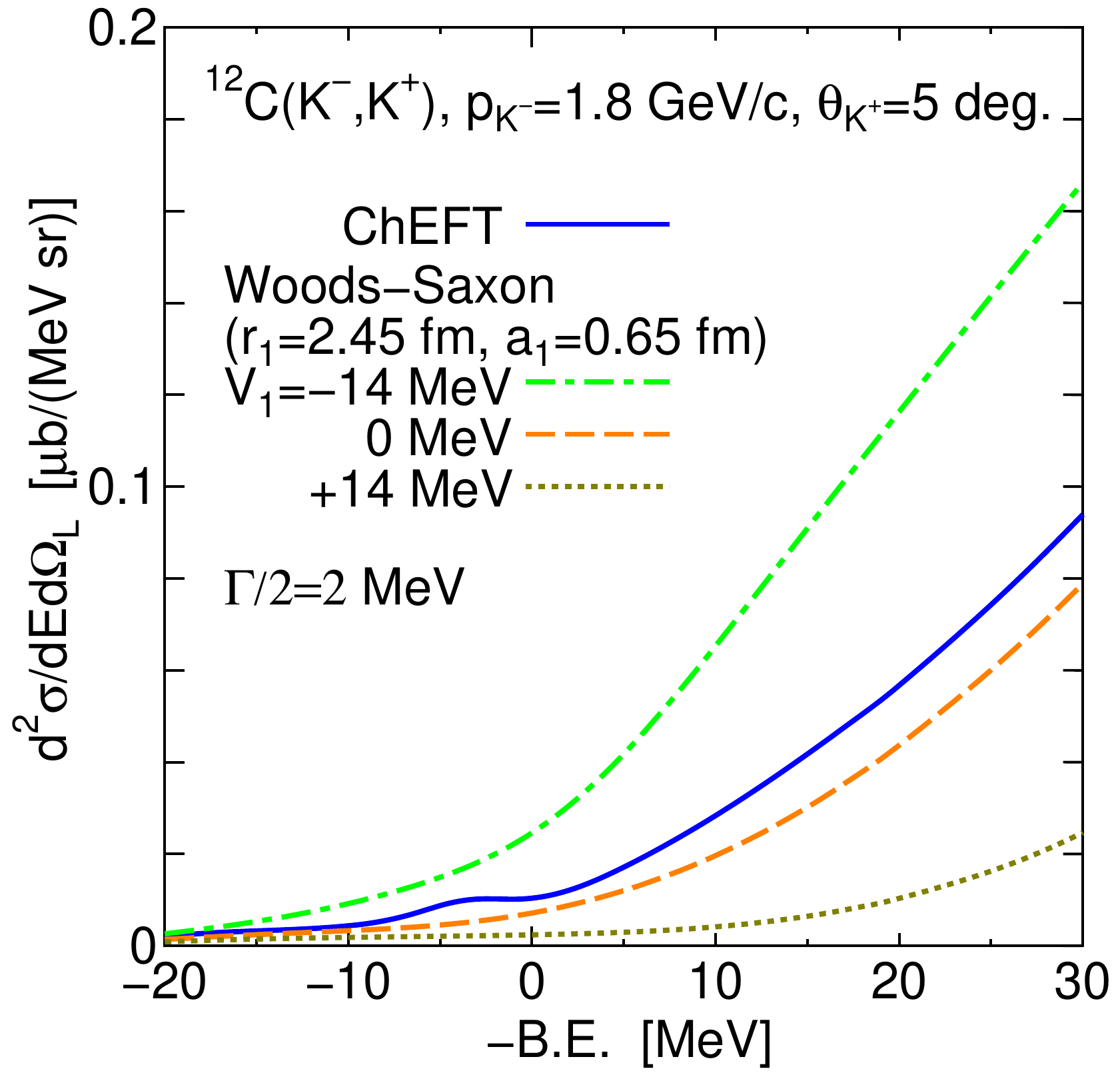}
 \caption{$K^+$ spectrum of $(K^-,K^+)$ $\Xi$ inclusive production
reactions on $^{12}$C around the threshold region with the incident $K^-$
momentum of $p_{K^-}=1.8$ GeV/c and the reaction angle in the laboratory
frame of $\theta_{K^+}=5^\circ$. The width of $\Gamma/2=2$ MeV is applied,
instead of 5 MeV in Fig. \ref{fig:c12}. The experimental resolution of $\Delta E-2$
MeV is assumed.}
\label{fig:c12th}
\end{figure}

The magnified figure around the threshold is given in Fig. \ref{fig:c12th}. In this
case, the half-width of $\Gamma/2=2$ MeV is employed because the $\Xi$ imaginary
potential is very small at low energies as seen in Fig. 1. The contributions of the shallow
bound states, which are given in Table II, are visible as a bump, although no sharp
peak structure is expected after the inclusion of the width and the resolution.
In the case of a single Woods-Saxon potential of $V_1=-14$ MeV, the bump is not
noticeable because of the rather large strength leaked from above the threshold.  
The differential cross section with the ChEFT potential is again similar to that obtained
with the null potential, which is consistent with the Khaustov data \cite{Kha00} as was
discussed in Ref. \cite{KH10}.

It is worthwhile to note that the overall shape of the preliminary spectrum from
the KEL-E05 experiments, which was presented in HYP2018 by Nagae \cite{Nag18}
as counts/2 MeV, is very close to the dotted curve in Figs. \ref{fig:c12} and \ref{fig:c12th}.
The sensitivity of the differential cross sections around the threshold on
the strength of the $\Xi$ potential, which is demonstrated in Fig. 9, indicates that
the experimental determination of the cross-section value is decisively important.

\section{Conclusions}
Properties of the $\Xi$ hyperon in the nuclear medium, which are described by the
baryon-baryon interactions in the strangeness $S=-2$ sector derived on the
next-to-leading-order level in chiral effective field theory \cite{HAID16}, are explored.
The updated parameters of the interactions \cite{HAID19} with the cutoff scale of
550 MeV are used in the present calculation. The $\Xi$ single-particle potential is
first evaluated in SNM, using the lowest-order Brueckner theory. In this framework,
high-momentum components of the baryon-baryon interactions are regularized by
solving the $G$-matrix equation together with incorporating the medium effects
from the Pauli exclusion principle and the change of the baryon propagator in the
nuclear medium. The $\Lambda$, $\Sigma$, and nucleon single-particle potentials
obtained in the preceding nuclear matter calculations \cite{Koh13,Koh17} with ChEFT
interactions are used for the corresponding baryon propagators in the baryon-channel
coupled $G$-matrix equations. The resulting $\Xi$ potential is weakly attractive, which
is caused by the baryon-channel coupling especially in the $T=1$ $^3S_1$ state.

The $\Xi$ potentials obtained in SNM are transformed to energy-dependent $\Xi$
potentials in finite nuclei by a local density approximation. The correction for finite
range effects is simulated by the convolution of a Gaussian form factor. Using the
potentials evaluated at $E=-5$ MeV and $E=0$ MeV, $\Xi^-$ bound states are calculated
for $^{12}$C and $^{14}$N. The obtained energies of the $0s$ and $0p$ states in $^{12}$C
and $^{14}$N are found to be rather close to experimental energies reported from
the emulsion experiments at KEK \cite{Aok09, Naka15}, corresponding to either
a $0s$ or a $0p$ state.

The shape of the potential obtained numerically can be parametrized as a sum of
an attractive part and a repulsive part both in a Woods-Saxon form. The strength
and geometry parameters of each part are moderately energy dependent. 
The parametrized $\Xi$ potentials in finite nuclei are applied to calculate $K^+$
spectra of the $(K^-, K^+)$ $\Xi$ production inclusive reactions on $^9$Be and
$^{12}$C, using a semiclassical distorted-wave method \cite{Hash08}. The comparison
of the results with the experimental data on $^9$Be shows that the absolute value of
double differential cross sections is properly reproduced, but the quasi-free peak
position locates at somewhat lower $\Xi$ energy than that of the experimental data,
which indicates that the $\Xi$ potential with a more repulsive strength is preferred.
Regarding the $K^+$ spectrum of the $(K^-, K^+)$ $\Xi$ production inclusive reaction
on $^{12}$C, the sensitivity of the double differential cross sections around the
threshold on the $\Xi$ potential is demonstrated. The quantitative cross-section
data of the $(K^-, K^+)$ experiments carried out at KEK will provide valuable
information on the character of the $\Xi$-nucleus potential.

There are considerable uncertainties at present in determining ChEFT coupling
constants of the baryon-baryon interactions in the strangeness $S=-2$ sector
because of the scarceness of experimental information. Considering that direct
$\Xi N$ scattering experiments are not feasible in the near future, the comparison
between the future experimental data of either $\Xi$ bound states in nuclei or $\Xi^-$
production reactions on nuclei with better accuracy than before and the theoretical
description for these observables through microscopic calculations is valuable to
reduce the uncertainties.  Recently, the derivation of baryon-baryon interactions
from calculations in the lattice formulation of quantum chromodynamics (LQCD)
has been progressing. It may be beneficial to compare the characters of the
ChEFT and the LQCD $\Xi N$ interactions \cite{Sas18} in each spin, isospin,
and angular momentum channel. 

\smallskip

{\it Acknowledgements.}
The author is grateful to J. Haidenbauer for providing him a code of chiral NLO
baryon-baryon interactions in the strangeness $S=-2$ sector and also for valuable
discussions. This work is supported by JSPS KAKENHI Grants No. JP16K17698 and No.
JP19K03849.

\end{document}